\documentclass[11pt,twoside]{article}
\usepackage{asp2014}

\aspSuppressVolSlug
\resetcounters

\begin{document}
\title{Radio Astronomy visibility data discovery and access using IVOA standards}
\author{Mireille~Louys,$^{1,2}$ Katharina~Lutz,$^2$ Yelena~Stein,$^2$  Anais~Egner,$^3$ and Fran\c cois~Bonnarel$^2$}

\affil{$^1$Universit\'e de Strasbourg, ICube, CNRS UMR 7357, Strasbourg, France \email{mireille.louys@unistra.fr}}
\affil{$^2$Universit\'e de Strasbourg, Observatoire Astronomique de Strasbourg, CNRS UMR 7550, Strasbourg, France}
\affil{$^3$Universit\'e de Strasbourg, IUT Robert Schumann, Dpt. Informatique, Strasbourg, France}

\paperauthor{Louys Mireille}{mireille.louys@unistra.fr}{0000-0002-4334-1142}{University of Strasbourg}{ICube UMR7357-CNRS, Observatoire Astronomique UMR7357-CNRS} {Strasbourg}{}{67000}{France}
\paperauthor{Bonnarel Fran\c{c}ois}{francois.bonnarel@astro.unistra.fr}{}{University of Strasbourg}{Observatoire Astronomique,  UMR7550-CNRS}{Strasbourg}{}{67000}{France}
\paperauthor{Katharina Lutz}{klutz@astro.unistra.fr}{0000-0002-6616-7554}{University of Strasbourg}{Observatoire Astronomique UMR7550-CNRS} {Strasbourg}{}{67000}{France}
\paperauthor{Yelena Stein}{yelena.stein@astro.unistra.fr}{}{University of Strasbourg}{ICube UMR7357-CNRS, Observatoire Astronomique UMR7550-CNRS} {Strasbourg}{}{67000}{France}
\paperauthor{Egner Anais}{anais.egner@live.fr}{}{University of Strasbourg}{IUT Robert Schumann, Dpt. Informatique}{Strasbourg-Illkirch}{}{67000}{France}

\begin{abstract} 
Enhancing interoperable data access to radio data has become a science priority within the International Virtual Observatory Alliance (IVOA). 
This lead to the foundation of the IVOA Radio astronomy Interest Group. Several radio astronomers and project scientists enrolled in various projects (NRAO, ASKAP, LOFAR, JIVE, ALMA, SKA, INAF, NenuFAR, etc.) have joined. Together they are paving the way to a better integration of their services in the virtual observatory (VO) infrastructure and propose extension of IVOA standards to help achieving this goal.

Calibrated radio datasets such as cubes, images, spectra and time series can already be searched and retrieved using the ObsCore/ObsTAP specification defined in the IVOA, or by dataproduct-specific services like SIAv2, SODA, SSA and ConeSearch. However, properties of radio visibility data are not fully implemented in the VO landscape yet. We need specific features to refine data discovery and selection that are adapted to radio astronomers' need.
In this context 
the VO team at the Centre de Donn\'{e}es astronomiques de Strasbourg (CDS) proposes to consider the ObsCore/ObsTAP specification \citep{2017ivoa.spec.0509L} and to establish cross-walks between the ObsCore and the existing Measurement Set (MS) metadata profile for data discovery of radio visibility data (VD). 

In order to account for the difference in granularity between radio VD datasets and science-ready datasets of the VO, the approach splits a MS data file into a list of datasets served by an ObsTAP service, thus enabling coarse grain discovery in the multi-wavelength context. 
Radio specific metadata such as number of antennae, frequency ranges, 
$uv$ plane coverage plots, frequency-phase and frequency-amplitude plots, primary and synthesized beams are also provided either by adding column metadata or by using the DataLink technique. Future evolution of this approach and lessons learnt are discussed.
\end{abstract}

\section{Goal }
Radio astronomical archives for large facilities (e.\,g. VLA, LOFAR, EVN, ATCA, ...) used to store mostly raw or calibrated visibility data. However, the situation is changing and some projects (e.\,g. ALMA, ASKAP, MeerKAT...), distribute science-ready data like cubes.
This approach is also planned for the SKA, the future Square Kilometer Array project.
Up to now raw visibility data are mostly accessed through project-specific web interfaces. Scientists can reduce the data with adjusted tools and customized configuration parameters. In the future, this workflow will evolve towards the use of next generation science platforms like
e.\,g. ESAP proposed within the ESCAPE\footnote{https://projectescape.eu/} project where reduction codes will run close to the data. The tremendous increase of observations provided by many large projects in the radio frequency domain is interesting beyond the radio astronomical community. Multi-messenger and multi-wavelength observations analysis is essential nowadays to explore and validate more advanced models of astrophysical processes involving complex interactions.

Many radio observatory archives already provide reduced data via VO-enabled services (e.g., CADC, CASDA, ATCA,...). These services include e.g. ObsTAP, HiPS servers, DataLink \citep{2015ivoa.spec.0617D} and SODA \citep{2017ivoa.spec.0517B}, for instance for science-ready radio cubes. 
In this study, we show how to adapt {ObsTAP} to serve as a VO interface for discovering and accessing especially visibility data from radio astronomy facilities.

\section{Metadata extraction from visibility-dedicated Measurement Set to ObsCore metadata profile}
The MS format \citep{2015A&C....12..174V} for radio visibility observations contains multiple exposures pointed on various sky regions which are stored in Fields.
Within each Field, the signal can be observed across several spectral windows and with various polarisation settings. The MS format wraps them together in one container. The CASA \citep{emonts2019casa} \emph{listobs} command displays a summary of the main metadata keywords of the observation. In addition  various plots and maps can be produced to interpret the information content such as $uv$ plane coverage maps, amplitude/phase plots or amplitude/$uv$ distance, etc.
In the ObsCore approach however,  multi-wavelength search requires feature separation in order to express how data are spanned on the spatial, spectral, temporal and observable axis.
Hence, data stored in a MS file must be represented as a collection of coverage-homogeneous subsets that ObsCore can describe.
We have set-up a split procedure for this. 
We consider one ObsCore dataset to be a subset of contiguous or overlapping SpectralWindows of same ChannelWidth for a given Field. 
Scans pointed towards the same Field are grouped together and gather all related time stamps for this dataset. 
This procedure has been trained on various test data stemming from ATCA, LOFAR, VLA, JIVE radio visibilities archives.
Fig .\ref{fig:workflow} illustrates the partition of an MS dataset as a collection of ObsCore member datasets.

\articlefigure[width=0.9\textwidth]{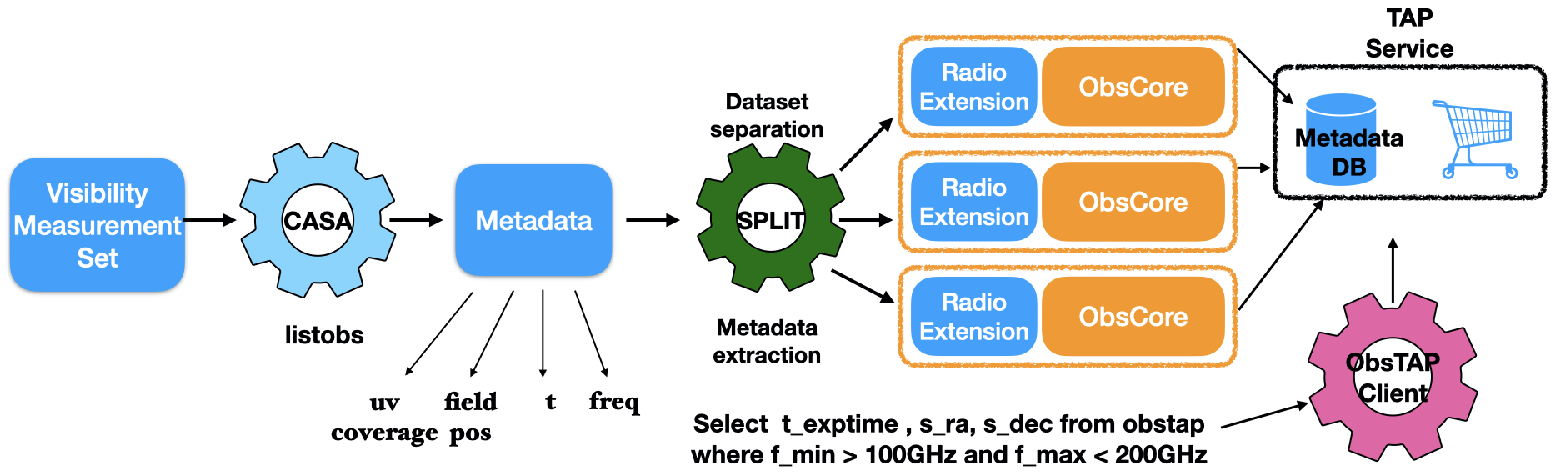}{fig:workflow}{The SPLIT procedure represents a MS dataset as a collection of members datasets whose coverage in spatial and spectral features in ObsCore is compact.}


Following the split procedure, an individual ObsCore dataset corresponds to one identified Field observed on one specific frequency interval, gathering contiguous channels. 
Such data set can be characterised using existing ObsCore keywords.
The \emph{data\-product\_type} is set to 'visibility'. The spatial features \emph{s\_ra}, \emph{s\_dec}, can be mapped to the reference sky position of one Field, \emph{s\_resolution} can be estimated from the longest baseline in the MS, and \emph{s\_fov} from the antennae diameter. Note that \emph{s\_fov} is dependent on the frequency. If the observations covers a large frequency range, one individual value for \emph{s\_fov} will only be a first approximation. 

The spectral coverage in radio 
observations
refers to frequency and multiple of Hz as units. 
We propose to add \emph{f\_min} and \emph{f\_max} for the frequency interval in kHz.  We would still keep the converted spectral band in \emph{em\_min}
and 
\emph{em\_max} in m as defined in ObsCore to allow multi-wavelength queries, and support interoperability between different ObsTAP services. 
In this case, the ObsCore convention allows to set \emph{em\_unit=kHz} and \emph{em\_ucd=em.freq}, for instance, in order to explicitly declare the usage of frequencies for the spectral information within the dataset.
The spectral resolution power \emph{em\_respower}, defined similarly in both wavelength and frequency also allows querying in both spectral quantities.

The VD values are complex Fourier coefficients stored in the
$uv$ plane; the observable axis of ObsCore can then represent it using \emph{o\_ucd=stat.Fourier}.
Time features \emph{t\_min}, \emph{t\_max}, \emph{t\_resolution} can directly be taken from the scan table within \emph{listobs} output. \emph{t\_exptime} is the sum of integration times on one Field and independent of the spectral coverage.

In radio astronomy the potential scientific interest for a dataset depends 
on the $uv$ plane coverage, sensitivity, signal-to-noise ratio, etc. The quality measures cannot be directly translated to one value, but instead are based on additional interpretation maps and instrument-dependent simulations. 
Software like CASA, for instance, allow to extract such maps and files from a MS dataset.

We have considered delivering these maps via a Datalink capability added to an ObsTAP service, as shown in Fig. \ref{fig:datalink}.
in order to provide access to :  1) the \emph{listobs} metadata summary file; 2) the MS file; 3) explanatory plots for 
the $uv$ plane coverage, 
antennae positions, amplitude/phase graphs, etc.

\articlefigure[width=0.9\textwidth]{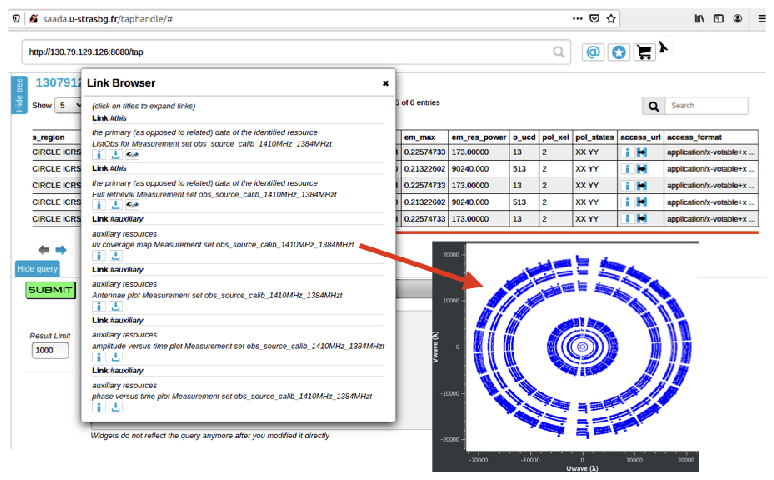}{fig:datalink}{ObsTAP query response with DataLink options to associated explanatory files. In blue the $uv$ plane coverage for a MS observation of NGC 104.}
A radio extension table for the ObsTAP standard is under study with LOFAR ASTRON and JIVE collaborators, to gather specific radio properties of these datasets:\\
  - s\_maxscale : the largest angular scale captured in the observation (LAS)\\  
  - n\_vis : the nb of visibilities covered within the individual dataset\\
  - $uv$ plane coverage filling factor\\
  - $uv$ plane coverage ellipse approximation or beam estimate (b\_max, b\_min, b\_angle)

\section{Conclusion}
This work demonstrates how to build a VO front-end for discovery of radio
visibility data, based on an ObsTAP service enriched with radio specific metadata selected here from the MS metadata profile. 

The extension of the ObsTAP TAP\_SCHEMA with an extra radio specific table is proposed for final discussion in the IVOA pages of the IVOA Radio Interest Group and will be consolidated using a wider set of radio datasets from various projects and data formats.

The effective partitioning of the data contained in the MS file into a set of data chunks corresponding to an individual ObsCore dataset is not required. 
This implies a duplication of the data volume which is undesirable for such amount of data. On the contrary, these data chunks can be generated on the fly by a customized 
extraction process that recombines or splits data chunks from the discovered MS datasets with the state-of-the-art tools appropriate to each particular radio archive. 
This approach would allow on the one hand full command of instrument/project specific processing to happen on the radio experts/ archives side and would, on the other hand, still facilitate multi-wavelength and multi-messenger data discovery for the astronomical community at large.

\acknowledgements For the support of ESCAPE (European Science Cluster of Astronomy and Particle Physics ESFRI Research Infrastructures) funded by the EU Horizon 2020 research and innovation program (Grant Agreement n.824064).

\bibliography{P9-72}  
\end{document}